\documentclass[a4paper,11pt]{article}

\usepackage{pos}
\usepackage{lipsum} 
\usepackage{subcaption}
\usepackage{wrapfig}
\usepackage{amsmath}
\usepackage{mathtools}
\usepackage{enumerate}
\usepackage{lineno}

\newcommand{\shrink}{\vspace{-0.3cm}}
\usepackage{titlesec}
\titlespacing*{\section}{0pt}{0.5\baselineskip}{0.6\baselineskip}

\usepackage{geometry}
\newlength{\bibitemsep}
\newlength{\bibparskip}
\let\oldthebibliography\thebibliography
\renewcommand\thebibliography[1]{%
  \oldthebibliography{#1}%
  \setlength{\parskip}{\bibitemsep}%
  \setlength{\itemsep}{\bibparskip}%
}

\title{A Two-Component Lateral Distribution Function for the Reconstruction of Air-Shower Events in IceTop}

\ShortTitle{Two-Component LDF}

\author{The IceCube Collaboration \\{\normalsize \normalfont(a complete list of authors can be found at the end of the proceedings)}\\}

\emailAdd{mark.weyrauch@kit.edu}
\emailAdd{dennis.soldin@kit.edu}

\abstract{The surface component of the IceCube Neutrino Observatory, IceTop, consists of an array of ice-Cherenkov tanks measuring the electromagnetic signal as well as low-energy ($\sim\rm{GeV}$) muons from cosmic-ray air showers. In addition, accompanying high-energy (above a few 100$\,\rm{GeV}$) muons can be observed in coincidence in the deep in-ice detector. A combined measurement of the low- and high-energy muon content is of particular interest for tests of hadronic interaction models as well as for cosmic-ray mass discrimination. However, since IceTop does not feature dedicated muon detectors, an estimation of the low-energy muon component of individual air showers is challenging. 

In this work, a two-component lateral distribution function (LDF), using separate descriptions for the electromagnetic and muon lateral distributions of the detector signals, is introduced as a new approach for the estimation of low-energy muons in air showers on an event-by-event basis. The principle of the air-shower reconstruction using the two-component LDF, as well as its reconstruction performance with respect to primary energy and number of low-energy muons will be discussed.

\vspace{4mm}
{\bfseries Corresponding authors:}
Mark Weyrauch$^{1*}$, Dennis Soldin$^{2,3}$\\
{$^{1}$\itshape Karlsruhe Institute of Technology, Institute for Astroparticle Physics, Karlsruhe, Germany}\\
\mbox{$^{2}$\itshape Karlsruhe Institute of Technology, Institute of Experimental Particle Physics, Karlsruhe, Germany}\\
{$^{3}$\itshape Department of Physics and Astronomy, University of Utah, Salt Lake City, UT 84112, USA}\\[4mm]

$^*$ Presenter

\ConferenceLogo{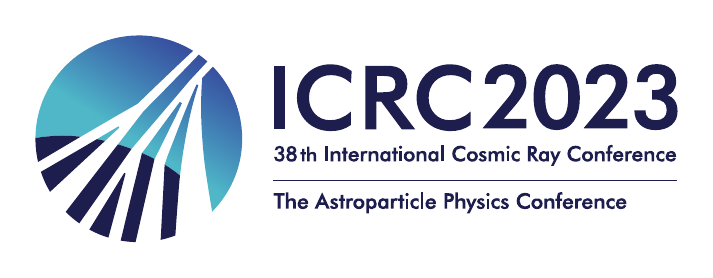}

\FullConference{The 38th International Cosmic Ray Conference (ICRC2023)\\ 26 July -- 3 August, 2023\\ Nagoya, Japan}
}

\begin{document}

\maketitle

\section{Introduction}\label{sec1}
Cosmic rays enter the Earth’s atmosphere where they produce extensive air showers (EASs)
which can be measured with large detector arrays at the ground. Although cosmic rays have been measured over several orders of magnitude in energy, large uncertainties remain, for example in measurements of their mass composition~\cite{Kampert:2012mx}. The main challenge lies in the understanding of muon production in EAS. Various experiments performed measurements of muons in EASs over the last decades, where some reported discrepancies in the number of muons in simulated and observed air showers, while others reported no discrepancies~\cite{Albrecht:2021cxw,EAS-MSU:2019kmv,Soldin:2021wyv,ICRC2023_WHISP}. IceTop has measured the GeV muon density in EASs at cosmic ray energies between $2.5\,\mathrm{PeV}$ and $120\,\mathrm{PeV}$ where no significant discrepancies have been observed~\cite{IceCubeCollaboration:2022tla}. However, this statistical analysis does not provide event-by-event information of the muon content in EASs which is important in order to correlate measurements of low-energy ($\sim$GeV) muons in IceTop and high-energy (above a few 100$\,\rm{GeV}$) muons in the deep-ice detector of the IceCube Neutrino Observatory~\cite{IceCube:2021ixw,Verpoest:2022ntf,ICRC2023_Stef}. This coincident measurement will provide strong constraints on hadronic interaction models~\cite{Coleman:2022abf,Soldin:2023wlh}. \\
In this work, we will present a novel approach to reconstruct the number of muons in air showers measured with IceTop on an event-by-event basis. This approach uses a two-component lateral distribution function (LDF) to fit the EAS signals measured in IceTop in order to account for the electromagnetic and muonic signal contributions separately.

\section{IceTop}\label{sec2}
IceTop~\cite{IceCube:2012nn} is the surface detector of the IceCube Neutrino Observatory (IceCube)~\cite{Aartsen:2016nxy} and it is located at the geographic South Pole at about $2.8\,\rm{km}$ above sea level, corresponding to an atmospheric depth of about $690\,\rm{g/cm}^2$. IceTop comprises $81$ detector stations, each consisting of two cylindrical
Cherenkov tanks, which are separated by approximately $10\,\rm{m}$, deployed in a triangular grid with a spacing of $125\,\rm{m}$. Each tank is filled with clear ice and houses two digital optical modules (DOMs) which measure the Cherenkov light generated by charged particles traversing the ice. An infill area in the center of the detector has a denser spacing of $< 50\,\rm{m}$, which improves the sensitivity of cosmic ray measurements at low energies. \\
A local trigger occurs when the signal in a tank exceeds a predefined discriminator threshold, as described in Ref.~\cite{IceCube:2012nn}. Further, a hard local coincidence (HLC) is defined where both tanks in a station have a local trigger within a time window of $1\,\mu\rm{s}$. If there is a local trigger in only one tank, it is called a soft local coincidence (SLC). Following calibration, which accounts for the individual tank responses, all IceTop signals are expressed in units of vertical equivalent muons~(VEM), which is the average signal produced by a muon traversing the tank vertically. For further analysis, various basic event cleanings are applied and only events that pass a cosmic ray filter are kept (see Ref.~\cite{IceCube:2012nn} for details). The subsequent EAS reconstruction is generally based on a fit to the lateral distribution of HLC hits (only) which provides the core position and direction of the air shower, as well as an estimate of its energy. In the following, we will describe an approach to extend this technique by including SLCs in order to reconstruct the lateral distribution of muons, which typically produce the SLC hits recorded by IceTop.
\newpage

\section{Signal Model}\label{sec3}
\begin{wrapfigure}{R}{0.6\textwidth}
\vspace{-15pt}
\shrink
  \begin{center}
    \includegraphics[scale=0.55]{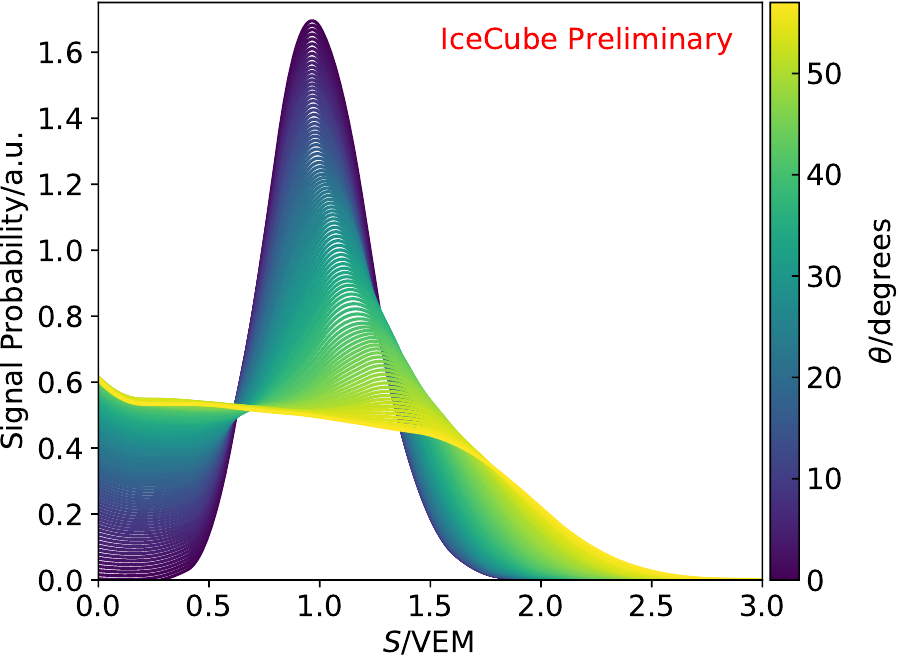}
  \end{center}
  \shrink
  \vspace{-9pt}
  \caption{Muon signal PDFs derived from IceTop tank response simulations for single muons injected at different zenith angles.}
  \label{fig:vert_muon_zenith_scan}
  \vspace{-12pt}
\end{wrapfigure}
In order to perform a combined fit of two LDFs a detailed understanding of the corresponding signal models is required. The signal response for the electromagnetic part of the shower is well-known from the standard IceTop reconstruction described in Ref.~\cite{IceCube:2012nn}. For the muonic part, the response of the IceTop tank is modeled via dedicated response simulations. Muons with different multiplicities and zenith angles are injected into the tank, followed by the simulation of the detector response using Geant4~\cite{GEANT4:2002zbu} and a calibration to units of VEM. 
Generally, the signal distributions of muons mainly depend on the track length within the detector, i.e. are determined by the tank geometry. The tank response for single muons as function of the zenith angle is shown in Fig.\ref{fig:vert_muon_zenith_scan}. 
The peak position for a vertical muon is by definition located at a charge deposit of 1\,VEM. With higher inclination the maximum track length increases, causing a shift in the peak position proportional to $1/\cos{\theta}$. Simultaneously, the influence of muons traversing only the edge of the detector manifests in an increased contribution at lower charge deposits. Hence, the width of the signal distribution, i.e. the signal fluctuation increases as a function of the zenith angle.
The signal distributions $p_{\mu,\rm{sig}}(S|\theta, n)$ for multiplicities of up to $n=15$ muons as function of muon inclination are saved as spline fits and can be retrieved for any given combination of observed signal strength and zenith angle. It is important to mention that the latter is given by the reconstructed primary zenith angle, assuming the muon directions follow the direction of the primary cosmic-ray.
In order to obtain the total muon signal probability density function (PDF) given an expectation value $\langle N_\mu \rangle$ for the average number of muons, the signal distributions for an integer number of muons have to be weighted by the according Poission probability
\begin{equation}
    p_\mu(S|\theta, \langle N_\mu \rangle) = \sum_{n}^{\infty} \frac{\langle N_\mu \rangle^n}{n!} e^{-\langle N_\mu \rangle}  p_{\mu,\rm{sig}}(S|\theta,n) \ , \label{eq:muPDF}
\end{equation}
where $\langle N_\mu \rangle$ is determined by the muon signal expectation value $\langle S_\mu \rangle$ given by the muon LDF multiplied with the effective IceTop tank area. For muon multiplicities above $n=15$ the muon PDF is approximated by a Gaussian distribution. The total PDF for an observed tank signal $S$ is determined by the convolution of electromagnetic (Gaussian) PDF
\begin{equation}
     p_{\rm{em}}(S|\theta,\langle S_{\rm{em}} \rangle) = \frac{1}{\sigma_{\rm{em},\log_{10}S}\sqrt{2\pi}} \exp \left(-\frac{\left(\log_{10}S-\log_{10}\langle S_{\rm{em}}\rangle\right)^2}{2\sigma^2_{\rm{em},\log_{10}S}} \right) \label{eq:emPDF}
\end{equation}
and the muon signal PDF. The electromagnetic signal is affected by the snow accumulation on top of the IceTop tanks. Thus, the resulting signal attenuation has to be taken into account in the reconstruction procedure.
\newpage


\section{Implementation}\label{sec4}
For the combined reconstruction of the electromagnetic and muonic part, respectively, two LDFs are fit to an observed charge distribution. The LDF used for the electromagnetic contribution is the Double Logarithmic Parabola (DLP) function~\cite{IceCube:2012nn}
\begin{equation}
    S_{\rm{em}} = S_{\rm{em},125} \left( \frac{r}{r_{\rm{em}}} \right) ^{-\beta_{\rm{em}}(S_{\rm{em},125}) - \kappa(S_{\rm{em},125}) \log_{10}{ \left( r/r_{\rm{em}} \right) }} , \ r_{\rm{em}} = 125\,\rm{m} \ . \label{eq:DLP}
\end{equation}
The parameters $\beta_{\rm{em}}$ and $\kappa$ describe the slope and the curvature of the function, respectively. $S_{\rm{em},125}$ is given as the signal strength at a distance of $125\,\rm{m}$ perpendicular to the shower axis and is used as a proxy for the primary energy. This reference distance is found to minimize the primary mass dependence while providing a good reconstruction resolution~\cite{Andeen:2011}. For the parametrization of $\kappa$ as a function of $S_{\rm{em},125}$, a separate simulation set is used in which all muons are omitted for the derivation of the detector response. The DLP function is then fit to the average lateral distributions in bins of $S_{\rm{em},125}$, resulting in both a decrease in the slope $\beta_{\rm{em}}$ and an increase in the curvature $\kappa$ (i.e., steeper tail of the distribution) for an accurate description of the observed lateral shape of purely electromagnetic showers. While the parametrization of $\kappa$ is fixed within the whole reconstruction procedure, the LDF slope is treated as free parameter with its parametrization taken as an initial choice. The LDF used to model the muon contribution is based on the Greisen function~\cite{Greisen:1960wc}:
\begin{equation}
    S_{\mu} = S_{\mu,600} \left(  \frac{r}{r_{\mu}} \right)^{-\beta_{\mu}(S_{\rm{em},125})} \left(  \frac{r+320\,\rm{m}}{r_{\mu}+320\,\rm{m}} \right)^{-\gamma(S_{\rm{em},125})} , \ r_\mu = 600\,\rm{m} \ , \label{eq:Greisen}
\end{equation}
containing two slope parameters $\beta_\mu$ and $\gamma$. Both are parametrized as a function of $S_{\rm{em},125}$, accounting for the steepening of the muon density profile with increasing primary energy. $S_{\mu,600}$ is used as a proxy for the low-energy muon number and is derived at a distance of $600\,\rm{m}$ in order to reduce the electromagnetic background while minimizing the influence of large fluctuations far from the shower axis. The distance of $320\,\rm{m}$ is the Greisen radius~\cite{Greisen:1960wc}. In order to constrain the parameters to the expected physical range and at the same time increase the stability of the reconstruction procedure, all parameters are fit while applying a penalty term
\begin{equation}
p_{\rm{constr}} = 1/2\sum_{p}\lambda_{p}(p-p_{\rm{par}})^2, \ p \in \{\log_{10}S_{\rm{em},125},\beta_{\rm{em}},\beta_\mu,\gamma\}, \ \lambda_p = \sigma_p^{-2} \label{eq:penalty}
\end{equation}
to the final likelihood (LLH) function (Eq.~\eqref{eq:tot_llh}). In the case of $S_{\rm{em},125}$, a constant penalty is applied, corresponding to the observed maximum spread of 4\% when omitting all muons from a standard IceTop reconstruction~\cite{IceCube:2012nn}. The effective strength $\lambda_p$ of the constraints for the slope parameters is determined by the inverse of the observed spread $\sigma_p^{-1}$ in the corresponding parameter distributions after application of the reconstruction procedure. This spread reduces as a function of $S_{\rm{em},125}$ as the reconstruction becomes more stable with increasing multiplicity. Thus, the corresponding parametrization of $\lambda_p$ results in an increased strength of the penalty with increasing $S_{\rm{em},125}$. In order to minimize a possible bias introduced by a given constraint, its parametrization is performed iteratively. With each iteration $p_{\rm{par}}$ and $\lambda_p$ are updated with the mean and spread of the parameter distributions, respectively. The reconstruction procedure itself is based on a negative log-likelihood minimization and is incorporated into a new reconstruction framework, designed for the reconstruction of multiple LDFs and/or detectors \cite{Lesz:2023icrc}. For the two-component fit, the combined PDF is utilized in the convolution regime to determine the likelihood value for a given tank signal:
\newpage
\setlength{\abovedisplayskip}{-0.4cm}
\begin{gather}
\begin{aligned}
   p \left( S|\theta , \langle S_{\rm{em}} \rangle, \langle N_{\mu} \rangle \right) = p_{\rm{trg}}
\begin{dcases}
    p_{\rm{em}}(\left(S-S_{\mu}\right)/c_{\rm{snow}}|\theta,\langle S_{\rm{em}} \rangle), \hspace{-8pt} & \sigma_{\mu} < \sigma_{\rm{em}}/10\\
    p_{\mu}(S-S_{\rm{em}}c_{\rm{snow}}|\theta,\langle N_{\mu} \rangle), \hspace{-8pt} & \sigma_{\rm{em}} < \sigma_{\mu}/10\\
    \int_{0}^{S} \hspace{-6pt} p_{\rm{em}}(S_{\rm{em}}'/c_{\rm{snow}}|\theta,\langle S_{\rm{em}} \rangle) p_{\mu}(S-S_{\rm{em}}'|\theta, \langle N_\mu \rangle) dS_{\rm{em}}', \hspace{-8pt} & \text{else} \ .
\end{dcases}
\end{aligned} \label{eq:p_signal}
\raisetag{15pt}
\end{gather}
To account for the snow accumulation, an exponential attenuation factor $c_{\rm{snow}}$~\cite{Rawlins:2023icrc} is applied to the electromagnetic signal. If either the electromagnetic or muon signal uncertainty (determined based on the corresponding signal expectation) is small in comparison, the according signal PDF is approximated as a delta function. The resulting signal PDF is multiplied with the discriminator trigger $p_{\rm{trg}}$~\cite{IceCubeCollaboration:2022tla} to account for the signal detection efficiency.
In addition to the resulting signal likelihood, saturated as well as silent (i.e. non-triggered) detectors are taken into account. While the former ($\rm{llh_{sat}}$) can be included as Gauss CDF~\cite{IceCube:2012nn}, the likelihood of silent detectors ($\rm{llh_{sil}}$) is based on the no-hit probability $p_{\rm{nohit}}$ derived from the convolution of the electromagnetic and muon contribution:
\setlength{\abovedisplayskip}{11pt}
\begin{equation}
   p_{\rm{nohit}} = p_{\rm{trg}}
\begin{dcases}
    \mathrm{Gauss CDF}((S_{\rm{thr}}-S_\mu)/c_{\rm{snow}}), & \sigma_{\mu} < \sigma_{\rm{em}}/10\\ 
    \vspace{-2pt}
    \sum_{n=0}^{3}  \frac{\langle N_\mu \rangle^n}{n!} e^{-\langle N_\mu \rangle} \int_{S_{\mu}=0}^{S_{\rm{thr}}-S_{\rm{em}}c_{\rm{snow}}}p_{\mu,\rm{sig}}, & \sigma_{\rm{em}} < \sigma_{\mu}/10\\
    \vspace{-2pt}
    \sum_{n=0}^{3}  \frac{\langle N_\mu \rangle^n}{n!} e^{-\langle N_\mu \rangle}\int_{S_{\mu}=0}^{S_{\rm{thr}}} \int_{S_{\rm{em}}=0}^{(S_{\rm{thr}}-S_\mu)/c_{\rm{snow}}}p_{\mu,\rm{sig}}p_{\rm{em}}, 
    \hspace{-6pt} & \text{else} \ .
\end{dcases}
\end{equation}
The convolution is calculated within similar integration regions as in Eq.~\eqref{eq:p_signal} and describes the probability of an observed tank signal to be below the threshold given a particular signal expectation from both the electromagnetic and muon contribution (as determined by the corresponding LDF). If both tanks in one station are not triggered (silent HLC), the likelihood is derived based on the squared probability $p_{\rm{nohit}}^2$:
\begin{equation}
   \rm{llh_{sil}} = \log\left( 1 - p_{\rm{nohit}}^n \right) , \ n=2/1 \ \text{if silent HLC/SLC} \ ,
\end{equation}
which corresponds to the not-hit probability of one station.
The global likelihood is derived by summation of all parts including the parameter constraints:
\begin{equation}
   \mathrm{llh} = \mathrm{llh_{sig}} + \mathrm{llh_{sat}} + \mathrm{llh_{sil}} - p_{\mathrm{constr}} \ . \label{eq:tot_llh}
\end{equation} 
The reconstruction procedure is realized as 6-step log-likelihood minimization. Starting out with a standard 3-step IceTop reconstruction as described in \cite{Lesz:2023icrc} as a basis for the two-component fit. In the last three steps, the two-component fit is performed while fixing the geometrical reconstruction derived from the previous steps.
Since the standard reconstruction only includes HLC hits, the underlying distribution as well as the resulting DLP fit are dominated by the electromagnetic contribution. Thus, the derived $S_{125}$~\cite{Aartsen:2013} parameter can serve as an anchor point for the electromagnetic function and determines the initial values of $S_{\rm{em},125}$ and all LDF slope parameters. Initially, $S_{\rm{em},125}$ is fixed and only the low-energy muon number estimator $S_{\mu,600}$ along with the muon and electromagnetic slope parameters are adjusted while applying the constraints shown in Eq.~\eqref{eq:penalty} in order to find the proper parameter space. In the last two steps, all LDF parameters except for $\kappa$ are fit successively. An example fit obtained after the last reconstruction step is illustrated in Fig.~\ref{fig:singleLDF_fit}, showing the electromagnetic and muon LDF as well as the resulting total LDF $S_{\rm{tot}}(r)=S_{\rm{em}}(r)+S_{\mu}(r)$.
\newpage
\setlength{\abovedisplayskip}{11pt}

\section{Reconstruction Performance}\label{sec5}
\begin{wrapfigure}{R}{9cm}
\shrink
\shrink
\vspace{5pt}
    \includegraphics[scale=0.6]{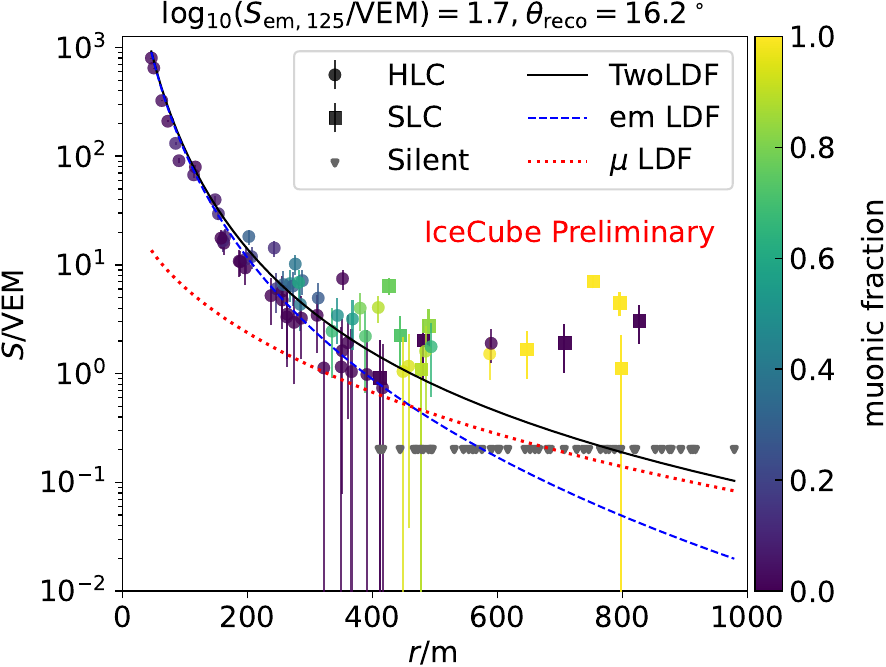}
  \shrink
  \caption{Two-component fit to the signal distribution of a simulated air-shower event with true primary energy of $\log_{10}E_{\rm{true}}=16.8\,\rm{eV}$. Silent detectors are visualized at a fixed value of $0.2\,\rm{VEM}$. The color scale indicates the fraction of the measured signal produced by muons as derived from CORSIKA.}
\vspace{-8pt}
\end{wrapfigure}
\begin{figure*}[ht!]
\shrink
    \centering
    \begin{subfigure}[h!]{0.48\textwidth}
    \centering
    \includegraphics[width=\textwidth,height=5cm]{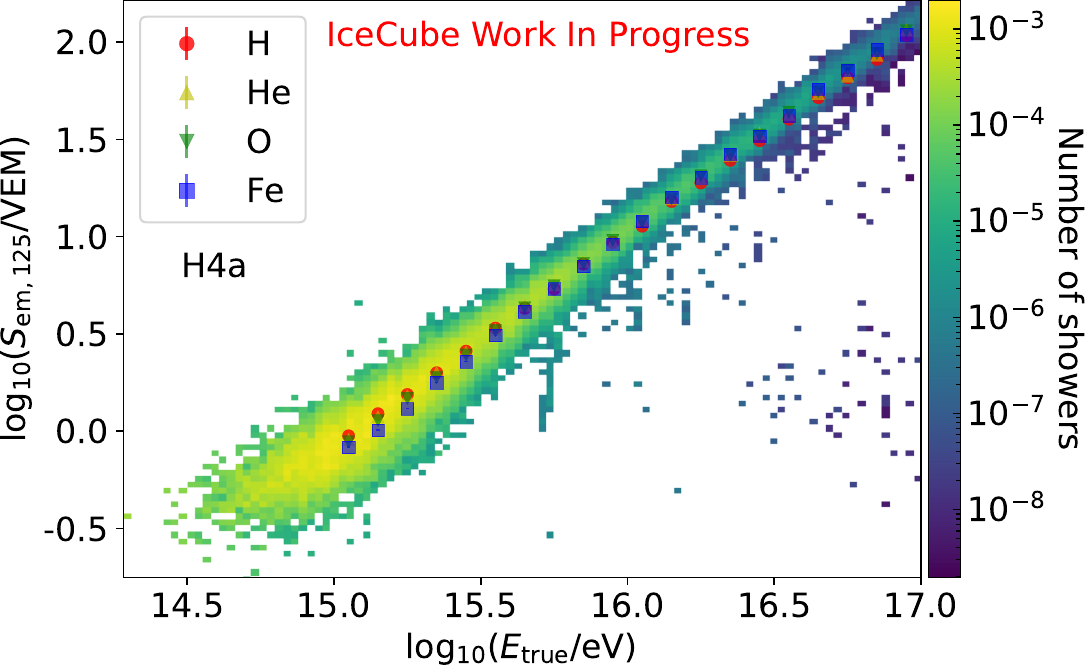}
    \end{subfigure}
    \hfill
    \begin{subfigure}[h!]{0.48\textwidth}
    \centering
    \includegraphics[width=\textwidth,height=5cm]{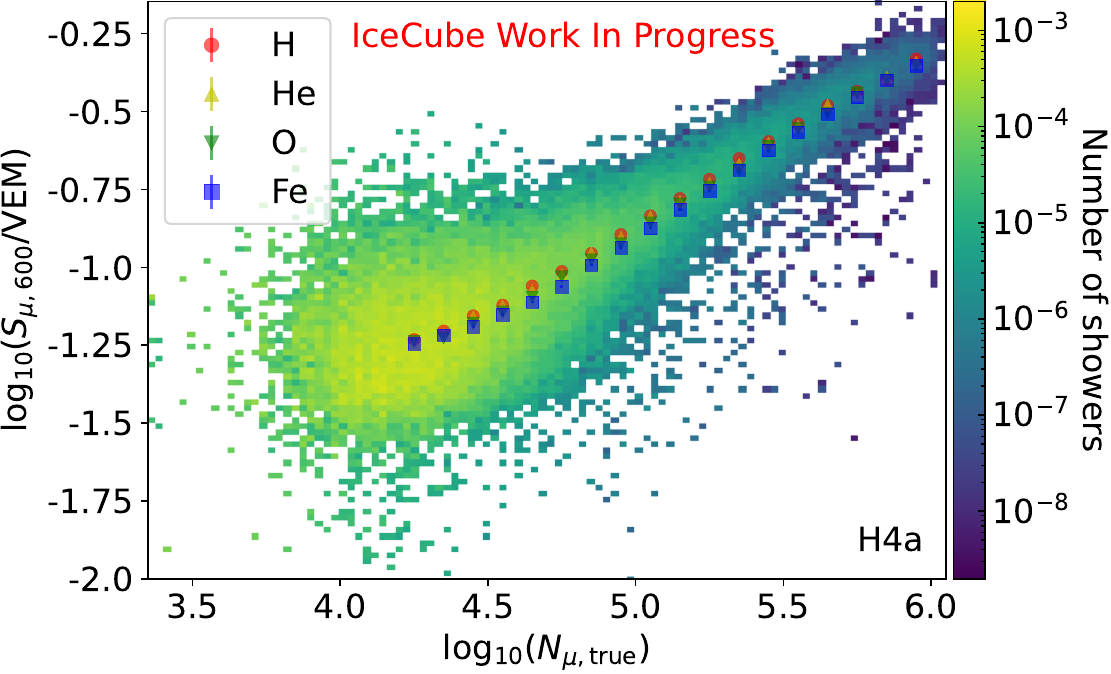}
    \end{subfigure}
    \shrink
    \caption{Distributions of the energy estimator $S_{\rm{em},125}$ (left) and the low-energy muon number estimator $S_{\mu,600}$ (right) for $\theta_{\rm{reco}}<18^\circ$ as a function of true primary energy and muon number, respectively, along with the mean values for each primary mass.}
    \label{fig:average_distributions}
\vspace{-0.08cm}
\end{figure*}

\begin{figure*}[ht!]
\shrink
    \centering
    \begin{subfigure}[h!]{0.482\textwidth}
    \centering
    \includegraphics[width=\textwidth]{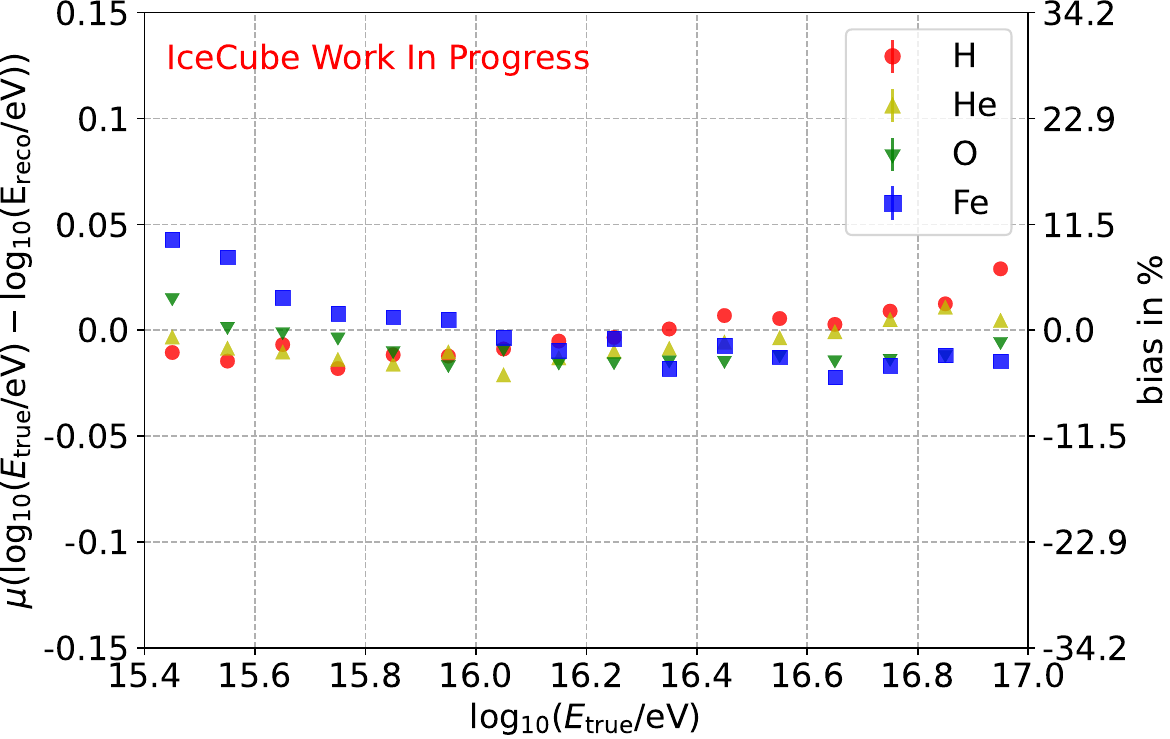}
    \end{subfigure}
    \hfill
    \begin{subfigure}[h!]{0.482\textwidth}
    \centering
    \includegraphics[width=\textwidth]{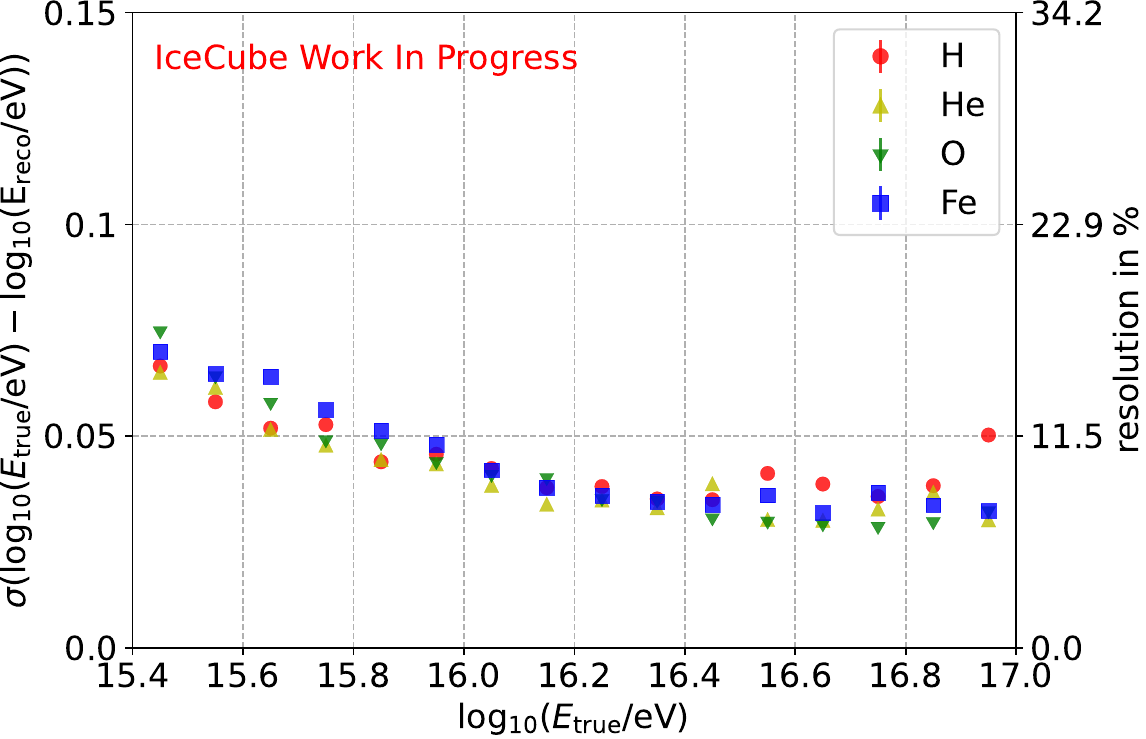}
    \end{subfigure}
    \begin{subfigure}[h!]{0.482\textwidth}
    \centering 
    \includegraphics[width=\textwidth]{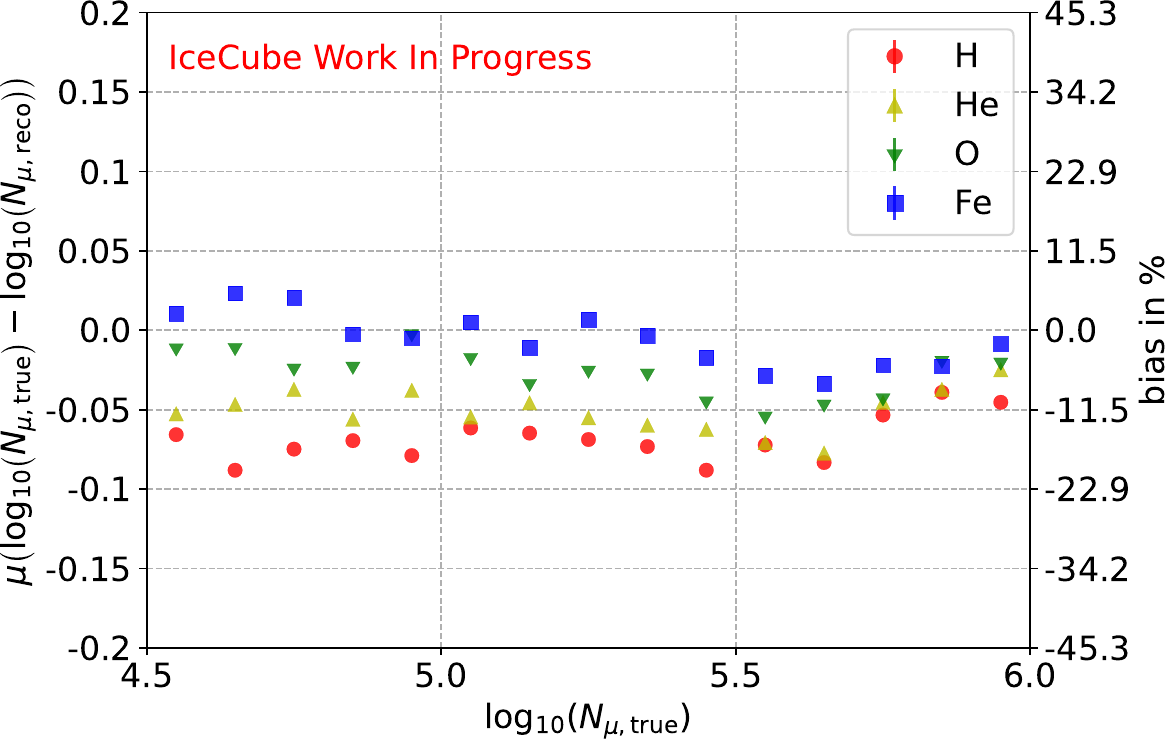}
    \end{subfigure}
    \hfill
    \hspace{8.5pt}
    \begin{subfigure}[h!]{0.482\textwidth}
    \centering 
    \includegraphics[width=\textwidth]{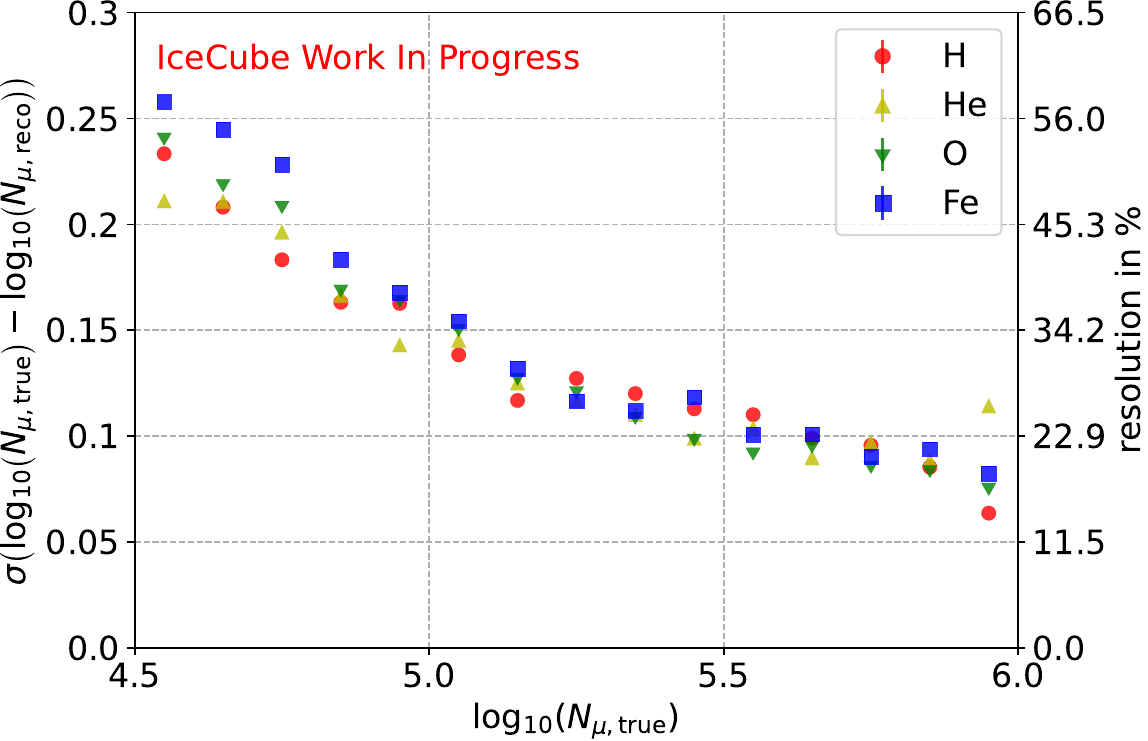}
    \end{subfigure}
    \shrink
    \caption{Reconstruction bias (left column) and resolution (right column) of the energy estimator $S_{\rm{em},125}$ (left) and the low-energy muon number estimator $S_{\mu,600}$ (right) as a function of true primary energy and muon number, respectively.}
    \label{fig:reco_performance}
\shrink
\end{figure*}

 \label{fig:singleLDF_fit}
The two-component reconstruction is applied to a simulation set produced with the Sibyll 2.1~\cite{Sibyll2.1:2009} hadronic interaction model, containing air showers of four different primaries (proton, helium, oxygen and iron) with energies of $100\,\rm{TeV}-100\,\rm{PeV}$. For the quantification of the reconstruction analysis, the application of different quality cuts is required. Firstly, a set of standard filters for the IceTop reconstruction discussed in~\cite{Aartsen:2013} is applied. Most importantly, this includes a containment cut with a simple geometrical requirement for the reconstructed shower core position to be within the IceTop array. Secondly, three additional cuts specific to the analysis presented in this work are applied. For a preliminary quantification of the reconstruction performance with the two component LDF only nearly vertical showers $\theta_{\rm{reco}}< 18^\circ$ are taken into account. To ensure a sufficient fit quality, only reconstructed fits resulting in a reduced $\chi^2$ of below 20 are included. This cut predominantly affects low energy events, as the fit stability increases with the event multiplicity. In the reconstruction only SLCs of $\geq 0.7\,\rm{VEM}$ are included in order to increase the probability of hits far from the shower axis to be of muonic origin~\cite{IceCubeCollaboration:2022tla}. To ensure a significant muonic contribution to the measured charge distribution, a final cut on the SLC multiplicity ($\geq 3$ SLCs) is applied. Above $\sim 20\,\rm{PeV}$ these cuts become practically independent of the primary mass. The resulting distribution for $S_{\rm{em},125}$ and $S_{\mu,600}$ is shown in Fig.~\ref{fig:average_distributions}, where $N_{\mu,\rm{true}}$ includes all muons above a threshold of 210\,MeV. Both distributions show a linear dependence to the corresponding true quantity, with a narrowing toward higher energies. The width of the distributions reflects the reconstruction resolution of the primary energy and low-energy muon number, respectively. 
In addition, the average of the distributions is shown for each simulated primary type separately. While $S_{\rm{em},125}$ as a function of the true energy shows only a small primary mass dependence, $S_{\mu,600}$ manifests a significant splitting between light and heavy primaries for the same $\log_{10}N_{\mu,\rm{true}}$. Although the mass dependence decreases with increasing muon number, it can result in a significant systematic shift for analyses utilizing $S_{\mu,600}$, and is therefore subject to further study. Fig.~\ref{fig:reco_performance} shows the reconstruction performance, i.e., the bias and resolution for both estimators. The primary energy is reconstructed with a minimal dependence (few percent level) on the cosmic-ray mass. Above 10\,PeV the energy resolution is below 10\% and improves toward higher energies. Around 100\,PeV saturation effects start to become important, and slightly worsen the reconstruction resolution for primary protons. 
The mass dependent bias of the low-energy muon number reconstruction is a direct reflection of the mass dependence of $S_{\mu,600}$. Towards higher $\log_{10}N_{\mu,\rm{true}}$ (i.e., higher primary energy), the bias reduces from 20\% to around 10\%. The corresponding reconstruction resolution improves as a function of $\log_{10}N_{\mu,\rm{true}}$ to below 20\%.

\section{Conclusion}\label{sec6}
In this work a new method for a combined reconstruction of the primary energy and the low-energy muon number of cosmic-ray air showers on an event-by-event basis is presented. For this purpose, a separate description for the electromagnetic and muon LDFs is used in a combined likelihood approach in order to fit the signal footprint measured with IceTop. It is shown, that the parameters $S_{\rm{em},125}$ and $S_{\mu,600}$ of the electromagnetic and muonic LDF, respectively, can be used as a proxy for the primary energy and the low-energy muon number. While the former can be reconstructed with a minimal dependence on the mass of the cosmic ray, the low-energy muon number estimation can be further optimized in terms of reconstruction resolution as well as primary mass dependence. Several possible reasons for the mass dependence in $S_{\mu,600}$, such as a non-optimal choice of the reference distance as well as a persisting entanglement between the electromagnetic and muonic part within the reconstruction procedure are currently under investigation. 
Further studies will include extended zenith and energy ranges as well as different levels of snow accumulation on top of the IceTop tanks. With increasing snow accumulation, the attenuation of the electromagnetic component becomes stronger, effectively increasing the potential of the IceTop tanks as muon detectors. This is of particular interest in the context of the IceTop enhancement~\cite{Haungs:2021} as well as the planned next generation detector IceCube-Gen2~\cite{Schroeder:2021}. 
An event-by-event based low-energy ($\sim$GeV) muon estimate is of high interest for tests of hadronic interaction models when combined with an estimation of the high-energy (above a few 100$\,\rm{GeV}$) muons of the same showers, allowing for a study of the corresponding correlations. Additionally, a separation of the electromagnetic and muonic shower component can be beneficial for mass composition analyses. Generally, any future study incorporating SLC information can potentially benefit from the application of the two-component LDF. In this context, the inclusion of IceTop uncontained events (shower core outside of IceTop) is of particular interest and will be part of future studies.

\bibliographystyle{ICRC}
\bibliography{references}

\clearpage

\section*{Full Author List: IceCube Collaboration}

\scriptsize
\noindent
R. Abbasi$^{17}$,
M. Ackermann$^{63}$,
J. Adams$^{18}$,
S. K. Agarwalla$^{40,\: 64}$,
J. A. Aguilar$^{12}$,
M. Ahlers$^{22}$,
J.M. Alameddine$^{23}$,
N. M. Amin$^{44}$,
K. Andeen$^{42}$,
G. Anton$^{26}$
% ...
\\
\\
% ...
$^{12}$ Universit{\'e} Libre de Bruxelles, Science Faculty CP230, B-1050 Brussels, Belgium \\
% ...
$^{17}$ Department of Physics, Loyola University Chicago, Chicago, IL 60660, USA \\
$^{18}$ Dept. of Physics and Astronomy, University of Canterbury, Private Bag 4800, Christchurch, New Zealand \\
% ...
$^{22}$ Niels Bohr Institute, University of Copenhagen, DK-2100 Copenhagen, Denmark \\
$^{23}$ Dept. of Physics, TU Dortmund University, D-44221 Dortmund, Germany \\
% ...
$^{26}$ Erlangen Centre for Astroparticle Physics, Friedrich-Alexander-Universit{\"a}t Erlangen-N{\"u}rnberg, D-91058 Erlangen, Germany \\
% ...
$^{40}$ Dept. of Physics and Wisconsin IceCube Particle Astrophysics Center, University of Wisconsin{\textendash}Madison, Madison, WI 53706, USA \\
% ...
$^{42}$ Department of Physics, Marquette University, Milwaukee, WI, 53201, USA \\
% ...
$^{44}$ Bartol Research Institute and Dept. of Physics and Astronomy, University of Delaware, Newark, DE 19716, USA \\
% ...
$^{63}$ Deutsches Elektronen-Synchrotron DESY, Platanenallee 6, 15738 Zeuthen, Germany  \\
$^{64}$ Institute of Physics, Sachivalaya Marg, Sainik School Post, Bhubaneswar 751005, India \\
% ...

\subsection*{Acknowledgements}

\noindent
The authors gratefully acknowledge the support from the following agencies and institutions:
USA {\textendash} U.S. National Science Foundation-Office of Polar Programs,
U.S. National Science Foundation-Physics Division,
U.S. National Science Foundation-EPSCoR,
Wisconsin Alumni Research Foundation,
Center for High Throughput Computing (CHTC) at the University of Wisconsin{\textendash}Madison,
Open Science Grid (OSG),
Advanced Cyberinfrastructure Coordination Ecosystem: Services {\&} Support (ACCESS),
Frontera computing project at the Texas Advanced Computing Center,
U.S. Department of Energy-National Energy Research Scientific Computing Center,
Particle astrophysics research computing center at the University of Maryland,
Institute for Cyber-Enabled Research at Michigan State University,
and Astroparticle physics computational facility at Marquette University;
Belgium {\textendash} Funds for Scientific Research (FRS-FNRS and FWO),
FWO Odysseus and Big Science programmes,
and Belgian Federal Science Policy Office (Belspo);
Germany {\textendash} Bundesministerium f{\"u}r Bildung und Forschung (BMBF),
Deutsche Forschungsgemeinschaft (DFG),
Helmholtz Alliance for Astroparticle Physics (HAP),
Initiative and Networking Fund of the Helmholtz Association,
Deutsches Elektronen Synchrotron (DESY),
and High Performance Computing cluster of the RWTH Aachen;
Sweden {\textendash} Swedish Research Council,
Swedish Polar Research Secretariat,
Swedish National Infrastructure for Computing (SNIC),
and Knut and Alice Wallenberg Foundation;
European Union {\textendash} EGI Advanced Computing for research;
Australia {\textendash} Australian Research Council;
Canada {\textendash} Natural Sciences and Engineering Research Council of Canada,
Calcul Qu{\'e}bec, Compute Ontario, Canada Foundation for Innovation, WestGrid, and Compute Canada;
Denmark {\textendash} Villum Fonden, Carlsberg Foundation, and European Commission;
New Zealand {\textendash} Marsden Fund;
Japan {\textendash} Japan Society for Promotion of Science (JSPS)
and Institute for Global Prominent Research (IGPR) of Chiba University;
Korea {\textendash} National Research Foundation of Korea (NRF);
Switzerland {\textendash} Swiss National Science Foundation (SNSF);
United Kingdom {\textendash} Department of Physics, University of Oxford.

\end{document}